\documentclass[12pt,aps,pre,preprint,showpacs,showkeys]{revtex4}

\usepackage{graphicx}
\usepackage{subfigure}
\usepackage{amsmath}
\usepackage{amssymb}
\usepackage{bm}

\begin{document}

\title{$XY$ model with antinematic interaction}
\author{Milan \v{Z}ukovi\v{c}}
 \email{milan.zukovic@upjs.sk}
 \affiliation{Institute of Physics, Faculty of Science, P. J. \v{S}af\'arik University, Park Angelinum 9, 041 54 Ko\v{s}ice, Slovakia}
\date{\today}

\begin{abstract}

We consider the $XY$ model with ferromagnetic (FM) and antinematic (AN) nearest-neighbor interactions on a square lattice for a varying interaction strength ratio. Besides the expected FM and AN quasi-long-range order (QLRO) phases we identify at low temperatures another peculiar canted ferromagnetic (CFM) QLRO phase, resulting from the competition between the collinear FM and non-collinear AN ordering tendencies. In the CFM phase neighboring spins that belong to different sublattices are canted by a non-universal (dependent on the interaction strength ratio) angle and the ordering is characterized by a fast-decaying power-law intra-sublattice correlation function. Compared to the FM phase, in the CFM phase correlations are significantly diminished by the presence of zero-energy domain walls due to the inherent degeneracy caused by the AN interactions. We present the phase diagram as a function of the interaction strength ratio and discuss the character of the respective phases as well as the transitions between them.

\end{abstract}





\maketitle

\section{Introduction}

A generalized ferromagnetic (FM) $XY$ model that includes a nematic term has been intensively studied in connection with various experimental realizations, such as superfluid A phase of $^3{\rm He}$~\cite{kors85}, liquid crystals~\cite{lee85,carp89,geng09}, or high-temperature cuprate superconductors~\cite{hlub08}. From a theoretical point of view such a model shows an interesting critical behavior with separate magnetic and nematic quasi-long-range order (QLRO) phases and the respective phase transitions belonging to different universality classes~\cite{lee85,kors85}. Recently, it has been revealed that the model, in which the nematic term is generalized to include higher-order (pseudo-nematic) couplings, can lead to a qualitatively different phase diagram with new ordered phases and phase transitions belonging to various universality classes~\cite{pode11,cano14,cano16}.

In the case of the model on a bipartite square lattice with a frustration parameter, it has been found that the phase diagram for the magnetic and nematic couplings of comparable strengths exhibits a phase in which the magnetism is ordered but the chirality remains disordered~\cite{qin09}. This phenomenon was ascribed to the competition between the two couplings in the formation of the chirality order. On the other hand, for a geometrically frustrated system on a non-bipartite triangular lattice with antiferromagnetic (AFM) and antinematic (AN) interactions the chiral long-range order has been confirmed in the absence of the magnetic order~\cite{park08}. 

The ground-state phase diagram of the frustrated classical Heisenberg and $XY$ models with biquadratic exchange interactions, was determined exactly considering a square- and a rhombic-symmetry versions~\cite{hayd10}. Recently, ground states of geometrically frustrated models with magnetic and generalized nematic couplings investigated in the exchange interactions parameter space have been shown to display a number of ordered and quasi-ordered phases as a result of geometrical frustration and/or competition between the magnetic and the generalized nematic interactions~\cite{zuko16}. Some of these models with the magnetic and nematic couplings having opposite signs have been proposed in the interdisciplinary applications for modeling of DNA packing~\cite{gras08} and structural phases of cyanide polymers~\cite{cair16,clark16,zuko16}. 

In the present study, we investigate the $XY$ model on a square lattice with the AN interactions. To our best knowledge, thermal and critical properties of this model have not been studied yet. We show that the competition between the FM and AN couplings leads to the change of the phase diagram topology featuring a new phase. Namely, besides the FM and AN QLRO phases, which are expected in the regions of dominance of the respective couplings, we identify at low temperatures another peculiar canted ferromagnetic (CFM) QLRO phase, which is wedged between the FM and AN phases. We focus on the character of the CFM phase, as well as phase transitions between the identified states.

\section{Model and methods}
The Hamiltonian of the generalized $XY$ model with the FM, $J_1$, and AN, $J_2$, interactions on a square lattice can be expressed in the form
\begin{equation}
\label{Hamiltonian}
{\mathcal H}=-J_1\sum_{\langle i,j \rangle}\cos(\phi_{i,j})-J_2\sum_{\langle i,j \rangle}\cos(2\phi_{i,j}),
\end{equation}
where $\phi_{i,j}=\phi_{i}-\phi_{j}$ is an angle between nearest-neighbor spins, $J_1 \equiv J \in (0,1)$ and $J_2 = J-1 < 0$. Notice that the opposite signs of $J_1$ and $J_2$ imply competition between the magnetic and nematic terms. While $J_1>0$ enforces a parallel spin alignment, i.e., $\phi_{i,j}=0$, $J_2<0$ prefers states with any perpendicular alignment of spins, with no preference for their orientation, i.e., $\phi_{i,j}=\pm \pi/2$.

Ground states of the model can be identified by finding global minima of the energy functional~(\ref{Hamiltonian}) in the phase space. Considering the fact that the square lattice is bipartite and assuming uniformity of spin ordering in the two interpenetrating sublattices, one basically needs to minimize the objective function
\begin{equation}
\label{OF}
f(\phi_{12})=-J_1\cos(\phi_{12})-J_2\cos(2\phi_{12}),
\end{equation}
where $\phi_{12}$ is the phase angle between the sublattices $1$ and $2$. This can be done analytically, however, particularly in complex cases like the present one with competing magnetic and nematic interactions, the sublattice uniformity assumption may not be justified. Therefore, care should be taken in order to find a true global minimum which, moreover, may not be unique. For that reason the analytical calculations are supplemented by numerical global optimization of the energy functional $\mathcal H$ on the entire lattice.


At finite temperatures we employ Monte Carlo (MC) simulations with Metropolis update. We consider spin systems of the sizes $L\times L$, with $L=24$-$120$, and apply periodic boundary conditions. Temperature dependencies of various thermodynamic quantities are obtained using standard MC simulation in which for thermal averaging we typically consider $2 \times 10^5$ MC sweeps (MCS) after discarding another $4 \times 10^4$ MCS necessary for thermalization (burn-in period). Randomly initialized simulations start from a high temperature (paramagnetic) region and gradually proceed to lower temperatures with a small step (typically $\Delta T=0.025$, measured in units of $J$ with the Boltzmann constant set to $k_{\rm B} \equiv 1$). The simulation at the next temperature starts from the final configuration obtained at the previous temperature. By following such a procedure one can achieve shortening of the thermalization period and make sure that the system remains close to the equilibrium during simulations in the entire temperature range.

If one is interested in the universality class of a given transition, it is useful to run much longer simulations close to the transition temperature and then apply reweighting techniques~\cite{ferr88,ferr89} for a certain range of the lattice sizes. This way one can more precisely localize maxima of various quantities involved in a finite-size scaling (FSS) analysis to determine the corresponding critical exponents. For that purpose we perform simulations using $10^7$ MCS after discarding initial $2 \times 10^6$ MCS for thermalization. Statistical errors are evaluated using the $\Gamma$-method~\cite{wolf04}.

The following thermodynamic functions are calculated, where $\langle\cdots\rangle$ denotes thermal averaging. The specific heat per spin $c$
\begin{equation}
c=\frac{\langle {\mathcal H}^{2} \rangle - \langle {\mathcal H} \rangle^{2}}{L^2T^{2}},
\label{c}
\end{equation}
magnetic $m_1$ and nematic $m_2$ order parameters
\begin{equation}
m_q=\langle M_q \rangle/L^2=\left\langle\Big|\sum_{j}\exp(\mathrm{i}q\phi_j)\Big|\right\rangle/L^2,\ q=1,2,
\label{m}
\end{equation}
and the corresponding susceptibilities $\chi_{q}$
\begin{equation}
\chi_{q} = \frac{\langle M_{q}^{2} \rangle - \langle M_{q} \rangle^{2}}{L^2T}.
\label{chi_q}
\end{equation}
We also evaluate a vortex density $\rho$, calculated directly from MC states. In particular, recalling that a vortex (antivortex) is a topological defect which corresponds to the spin angle change by $2\pi$ $(-2\pi)$ going around a closed contour enclosing the excitation core, they can be identified by summation of the angles between adjacent four spins on each square plaquette for each equilibrium configuration. The latter can result in $2\pi$ (vortex), $-2\pi$ (antivortex) or $0$ (no topological defect). Then the equilibrium defect density $\rho$ is obtained as a normalized thermodynamic average of the absolute value.

Further, we calculate the first and second rank correlation functions
\begin{equation}
\label{CF}
g_q(r) = \langle \cos(q\phi_{i,j}) \rangle,\ q=1,2,
\end{equation}
where $i$ and $j$ are two spins separated by a distance $r$, directly from MC states. Due to high computational complexity (the CPU time increases quadratically with the number of spins, i.e., $O(L^4)$ operation), we restricted our calculations to a moderate size of $L=48$ and considered the lags corresponding to the first $L/2=24$ nearest-neighbor distances. To obtain error estimates the values were averaged over $N_r = 20$ independent replicas. In the algebraic (BKT) phase the correlation function decays with the distance as a power law and, therefore, the exponent $\eta$ can be obtained from the relation
\begin{equation}
\label{g_FSS}
g_q(r) \propto r^{-\eta}.
\end{equation}
On the other hand, if the transition belongs to the Ising universality class, then maxima of the susceptibility should grow with the lattice size according to the FSS law 
\begin{equation}
\label{fss_chi}\chi_{q,max}(L) \propto L^{\gamma/\nu},
\end{equation}
where $\gamma$ and $\nu$ are critical exponents of the susceptibility and the correlation length, respectively.

\section{Results}
\subsection{Ground state}

\begin{figure}[t!]
\centering
\subfigure{\includegraphics[scale=0.48,clip]{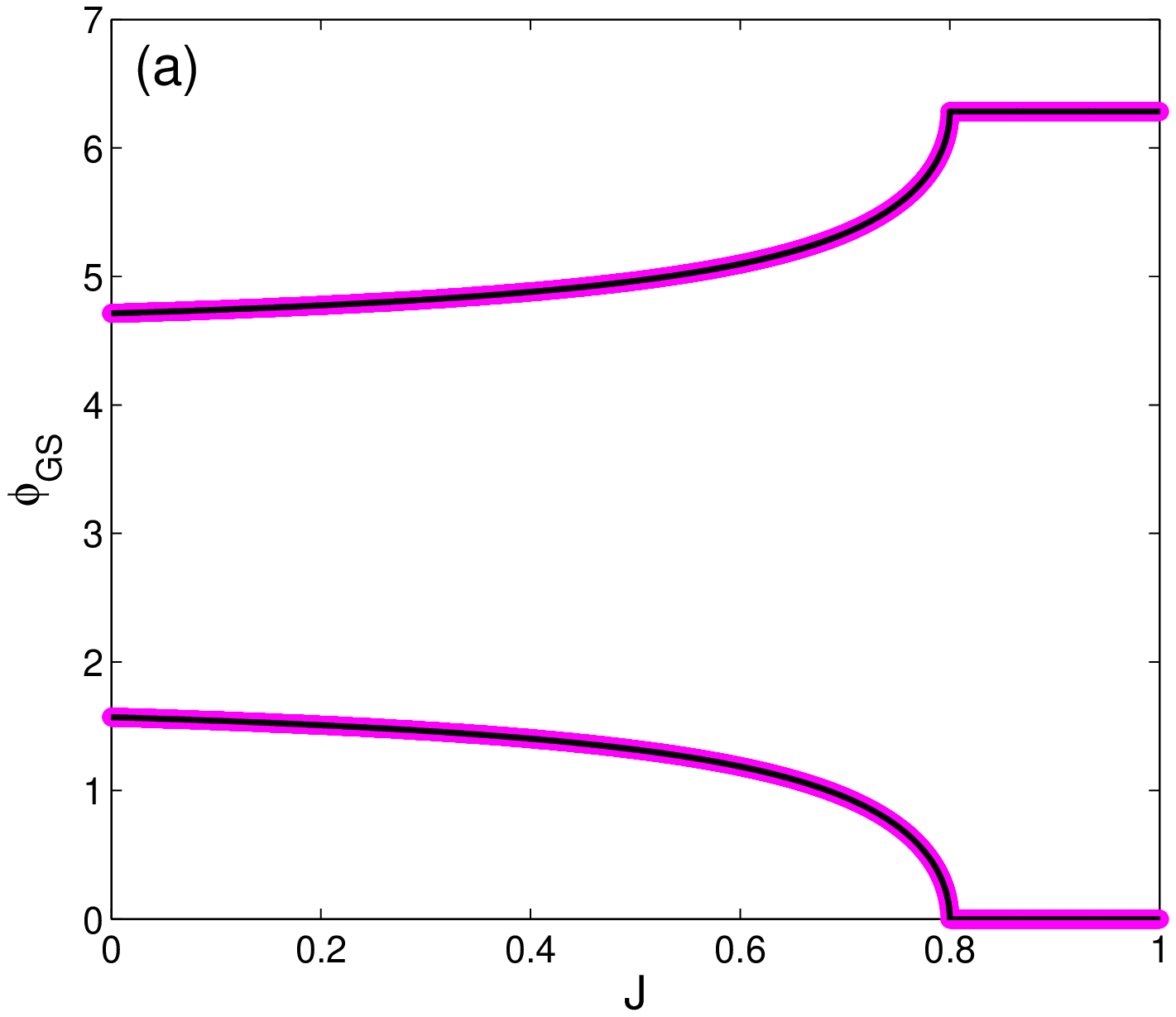}\label{fig:phi_gs}}
\subfigure{\includegraphics[scale=0.48,clip]{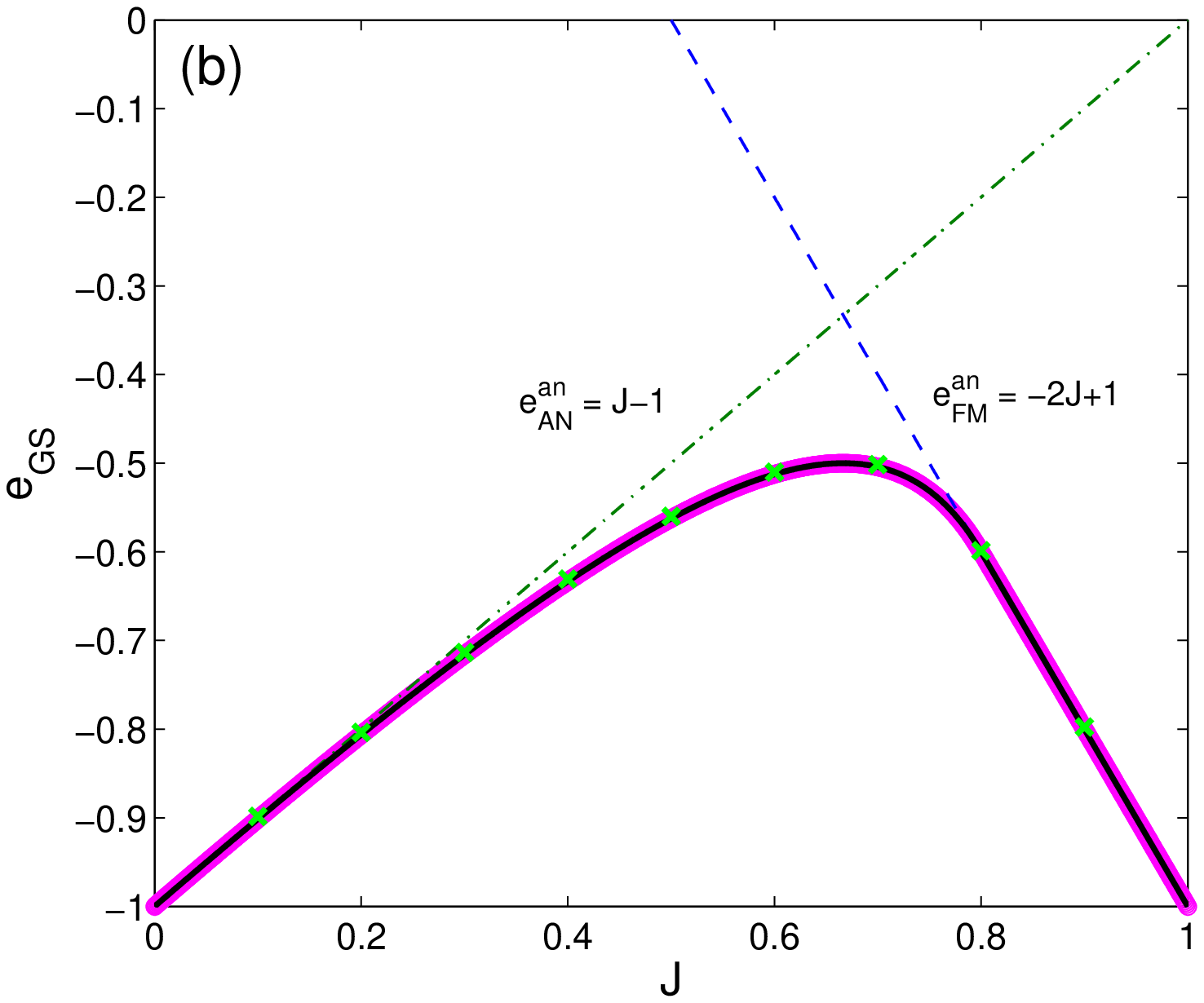}\label{fig:e_gs}}
\caption{(Color online) (a) Ground-state spin angles and (b) the corresponding energies per spin pair, obtained analytically (black solid curves) and from numerical optimization (magenta symbols). The dashed lines in (b) correspond to unstable solutions for the FM and AN states, within $J \in (0,0.8)$ and the green crosses the energy per spin pair obtained from MC simulations at the lowest considered temperature.}\label{fig:gs}
\end{figure}

\begin{figure}[t!]
\centering
\subfigure{\includegraphics[scale=0.56,clip]{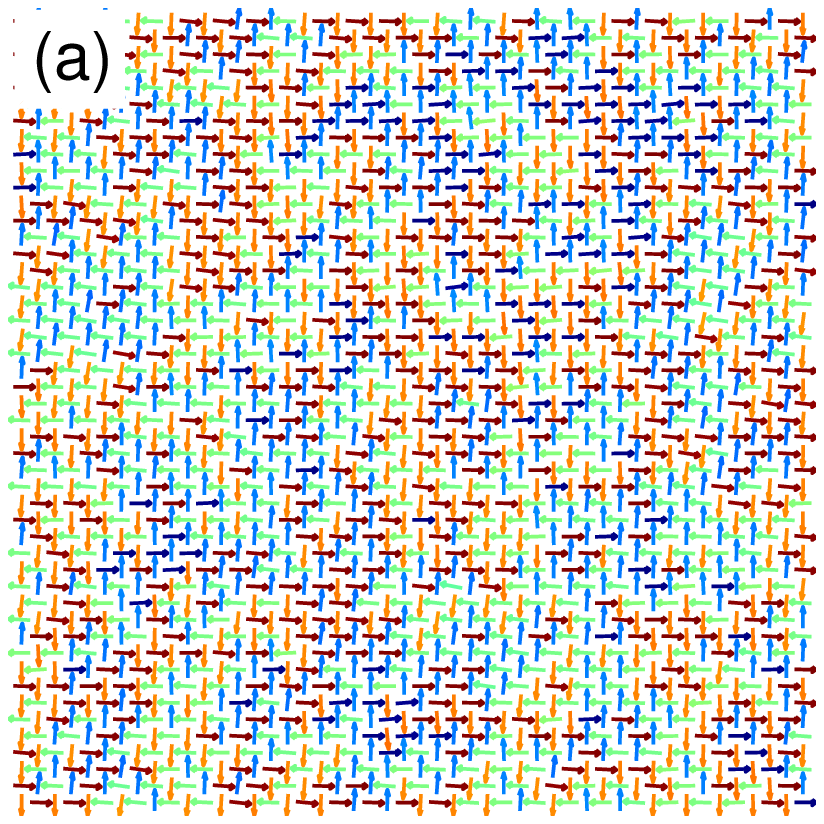}\label{fig:snap_J01}}
\subfigure{\includegraphics[scale=0.56,clip]{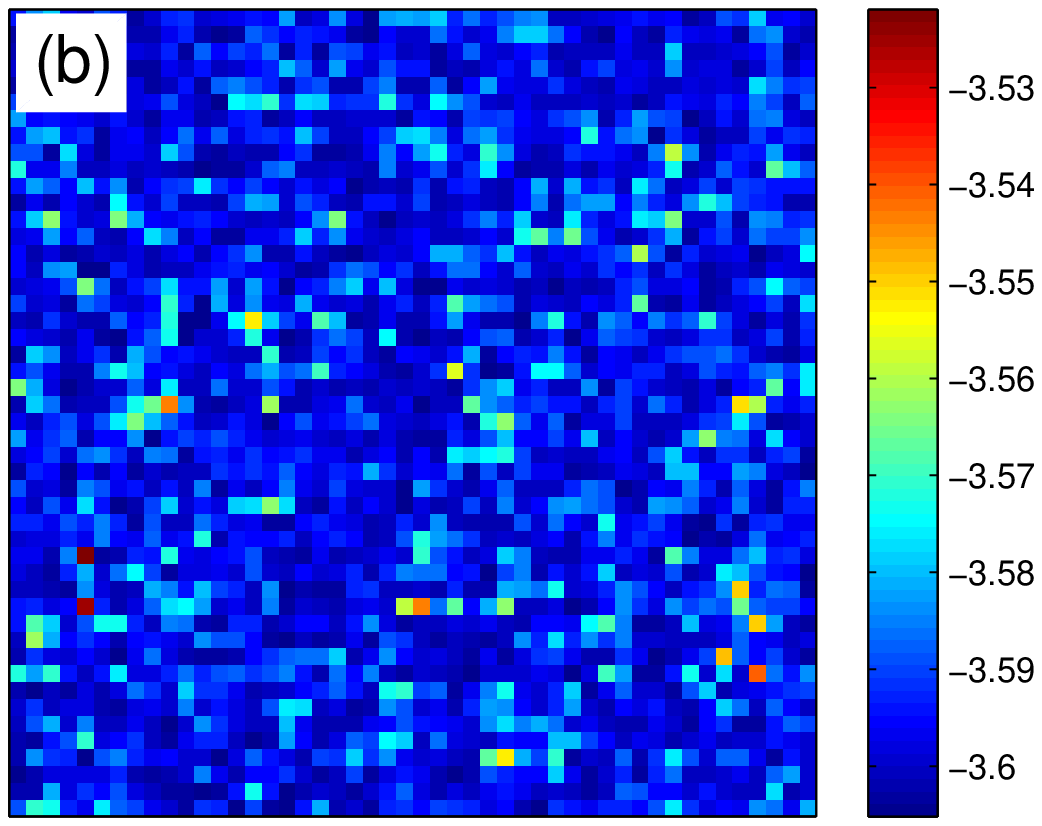}\label{fig:e_snap_J01}}\\
\subfigure{\includegraphics[scale=0.56,clip]{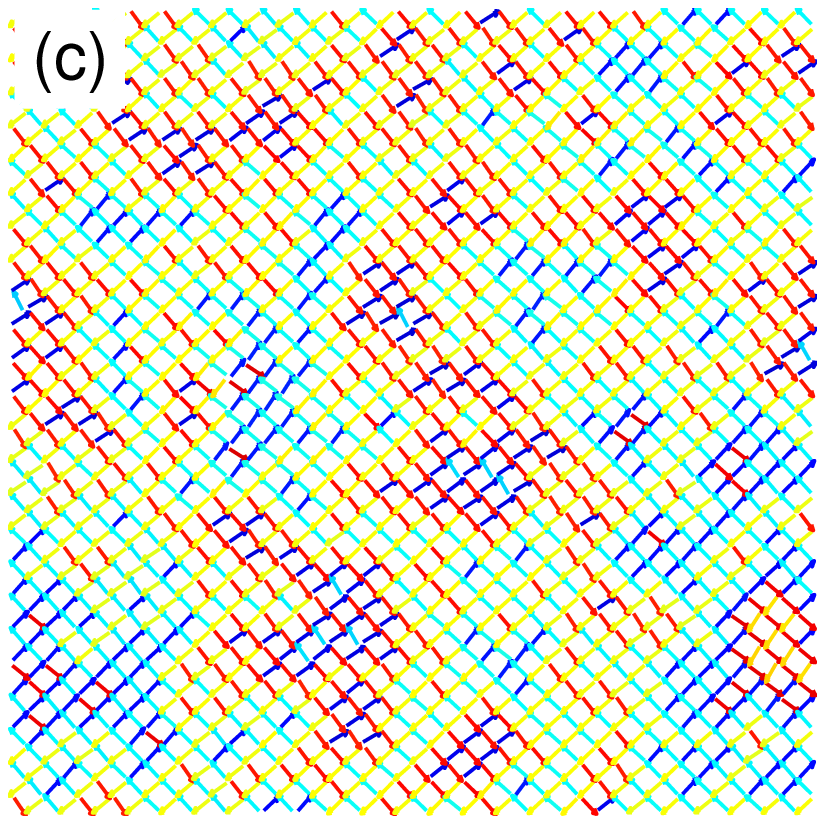}\label{fig:snap_J02}}
\subfigure{\includegraphics[scale=0.56,clip]{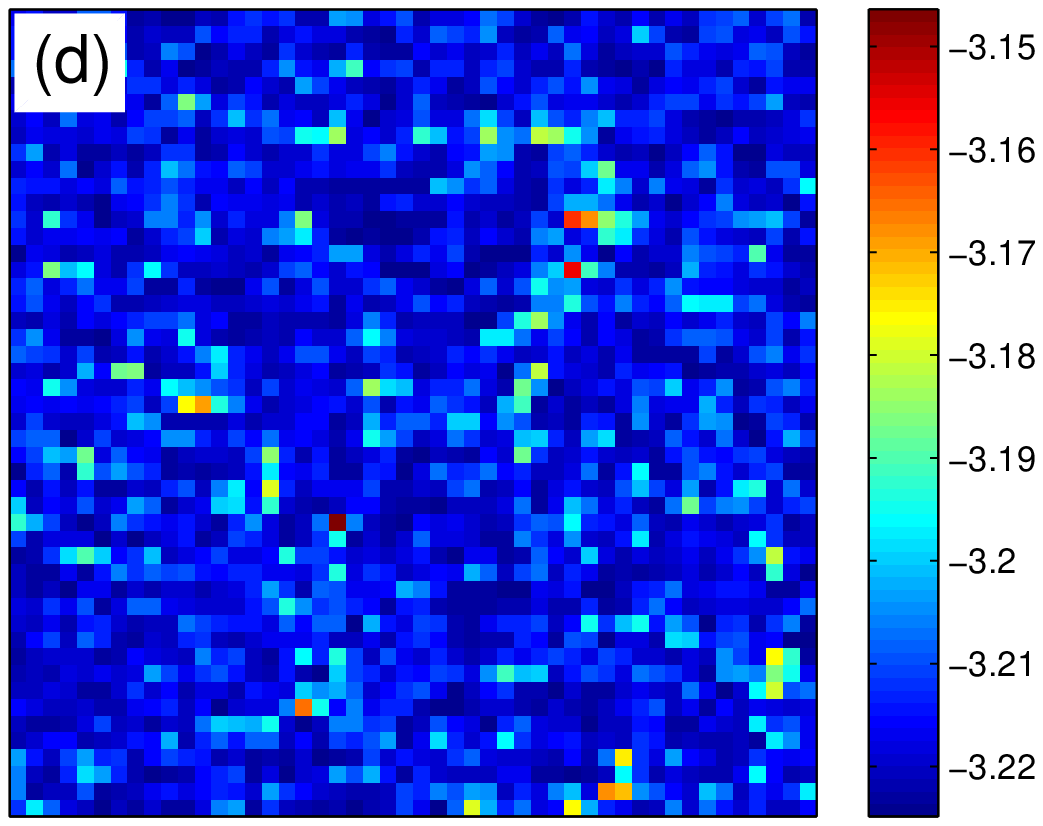}\label{fig:e_snap_J02}}\\
\subfigure{\includegraphics[scale=0.56,clip]{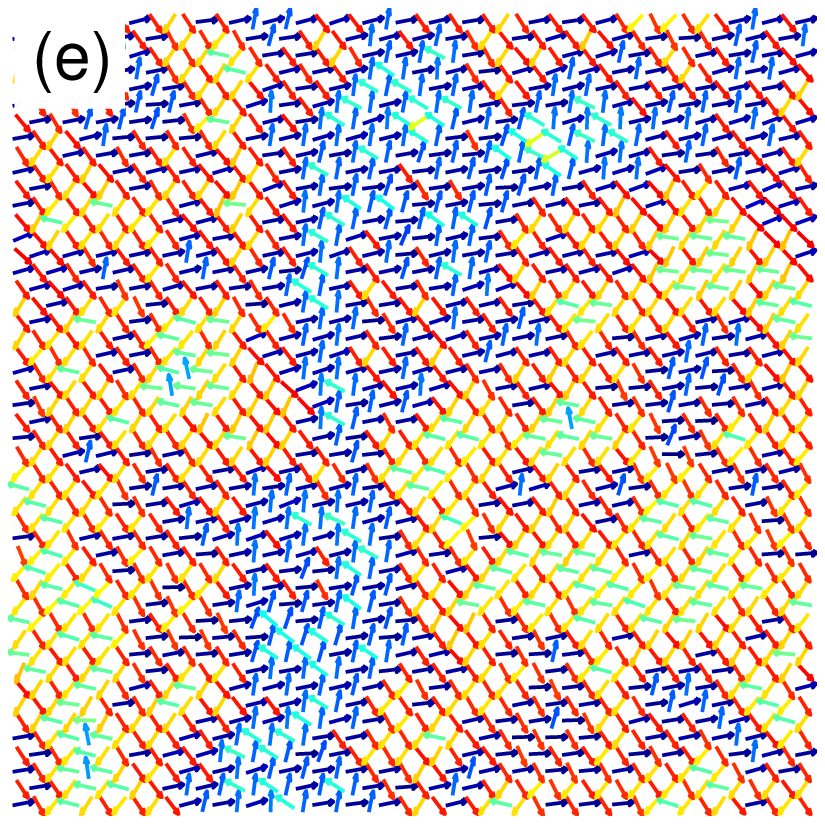}\label{fig:snap_J06}}
\subfigure{\includegraphics[scale=0.56,clip]{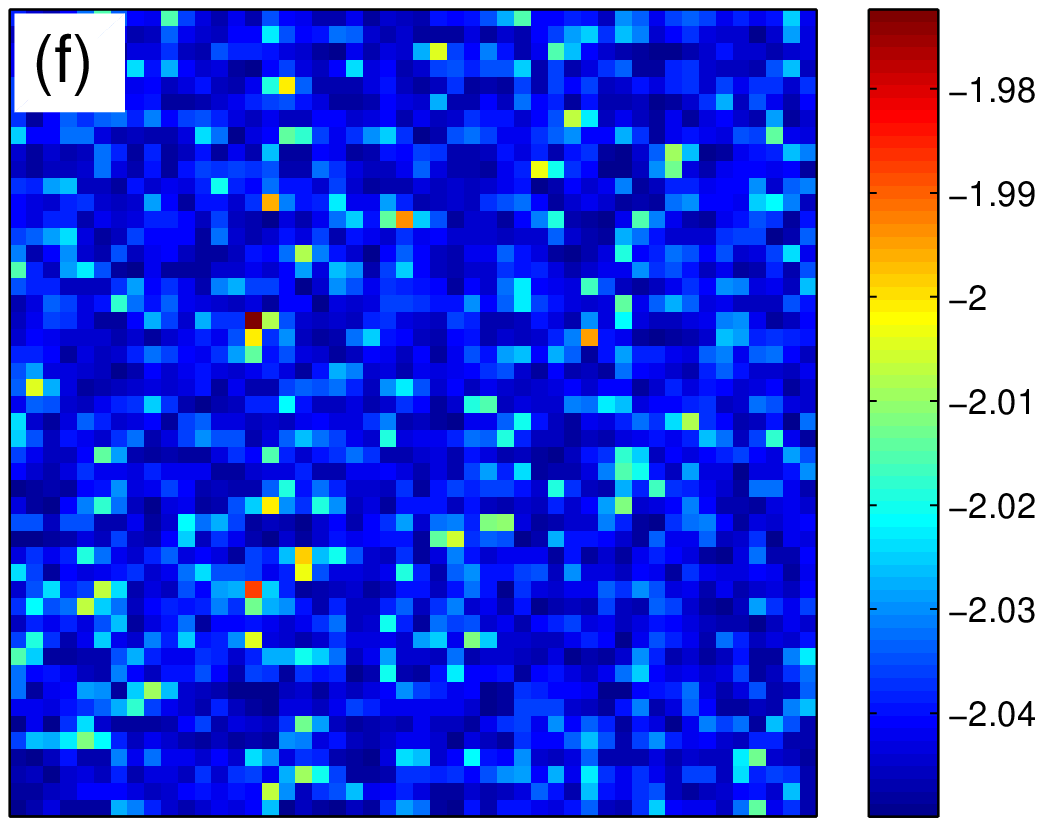}\label{fig:e_snap_J06}}\\
\subfigure{\includegraphics[scale=0.56,clip]{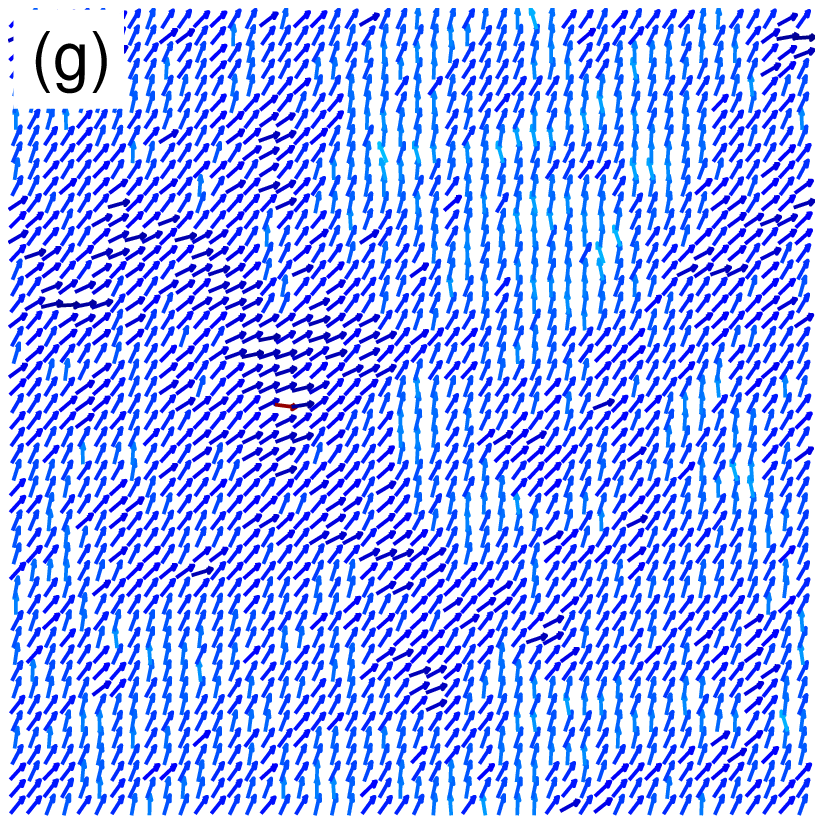}\label{fig:snap_J08}}
\subfigure{\includegraphics[scale=0.56,clip]{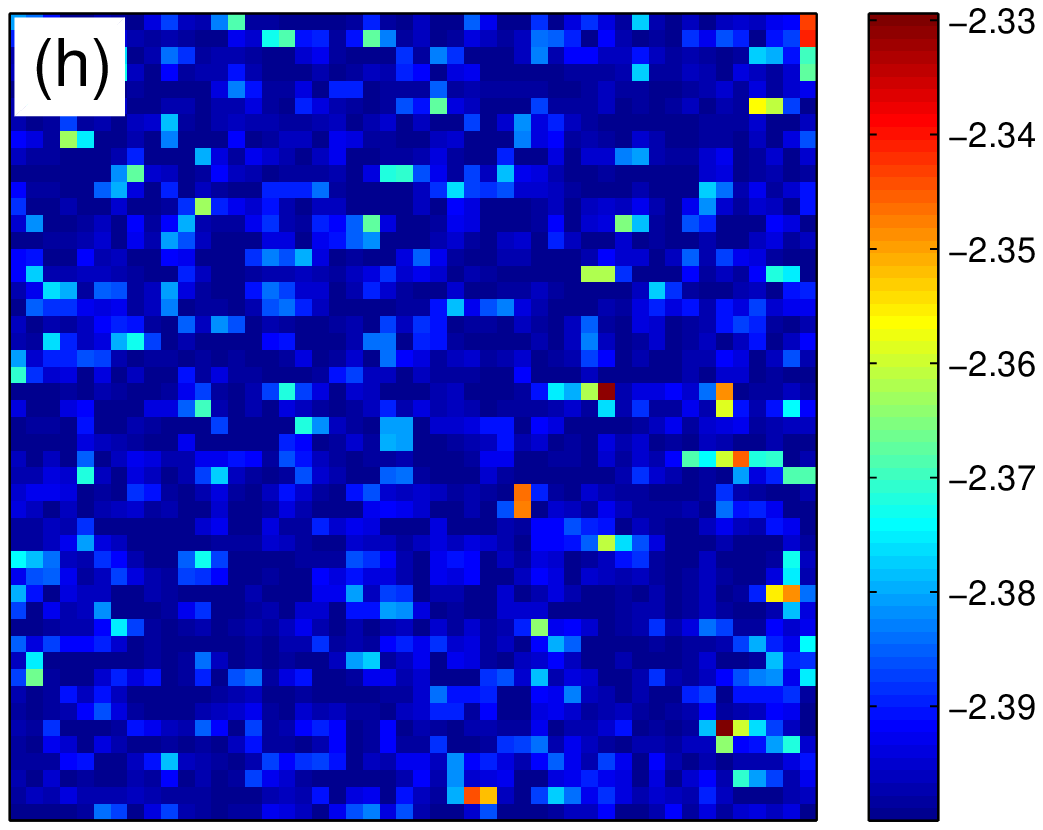}\label{fig:e_snap_J08}}
\caption{(Color online) Examples of spin configurations (left column) and local energy distributions (right column) in the CFM phase at $T=0.01$, for (a,b) $J=0.1$, (c,d) $J=0.2$, (e,f) $J=0.6$, and (g,h) $J=0.8$.}\label{fig:gs_snaps}
\end{figure}

Fig.~\ref{fig:phi_gs} shows ground-state spin angles obtained analytically in the form $\phi_{GS}^{an}=\arccos(-J/(4(J-1)))$, for $J \in (0,0.8)$, and $0$, for $J \in [0.8,1)$, (black solid curves) which are corroborated by the values $\phi_{GS}^{op}$ obtained from the numerical optimization (magenta symbols). Fig.~\ref{fig:e_gs} compares the energies per spin pair of different states. One can see, that within $J \in (0,0.8)$ the noncollinear states shown in Fig.~\ref{fig:phi_gs} with the energies $e_{GS}^{an}=-J\cos(\phi_{GS}^{an})-(J-1)\cos(2\phi_{GS}^{an})$ (black solid curve), evaluated analytically, or $e_{GS}^{op}$ (magenta symbols), calculated numerically, correspond to stable solutions. Hereafter we will refer to this phase as the canted ferromagnetic (CFM) phase. On the other hand, the ferromagnetic (FM) and antinematic (AN) states with perfectly parallel or perpendicular angles and the energies $e_{FM}^{an}=-2J+1$ and $e_{AN}^{an}=J-1$ (dashed lines), respectively, correspond to unstable solutions. Within $J \in [0.8,1)$ the ground state becomes the state with $\phi_{GS}=0$, i.e. the FM state.

For $J\equiv J_1=0$ no magnetic ordering can be expected due to the macroscopic degeneracy resulting from the two-fold degeneracy $\phi_{i,j}=\pm \pi/2$ of each nearest-neighbor spin pair. For $J > 0$ the local two-fold degeneracy $\phi_{i,j}=\pm \phi_{GS}$, where $|\phi_{GS}| < \pi/2$ persists. Nevertheless, in the snapshots taken from MC simulations close to zero temperatures (see the left column in Fig.~\ref{fig:gs_snaps}) one can observe formation of small domains of similarly oriented spins within each of the two sublattices of the square lattice the size of which gradually increases with the increasing value of the ferromagnetic coupling $J$. Eventually, for $J \geq 0.8$ the canting angle $\phi_{GS}$ becomes zero and the domains merge to a single ferromagnetic domain spanning the entire lattice. In the right column in Fig.~\ref{fig:gs_snaps} we present local energy distributions corresponding to the snapshots to their left. In all the instances there are only small fluctuations around the mean values, resulting from low but non-zero temperature $T=0.01$, with no traces of any boundaries separating different spin domains. These energy snapshots demonstrate that the spin domains present within the CFM phase have zero-energy walls due to the inherent degeneracy caused by the AN interactions. We note that at low-temperatures canted magnetic phases, resulting from the competition between the magnetic and nematic couplings, have also been reported in frustrated models with triangular geometry~\cite{zuko02,zuko03a,zuko03b}.

\subsection{Finite temperatures}

\begin{figure}[t!]
\centering
\subfigure{\includegraphics[scale=0.48,clip]{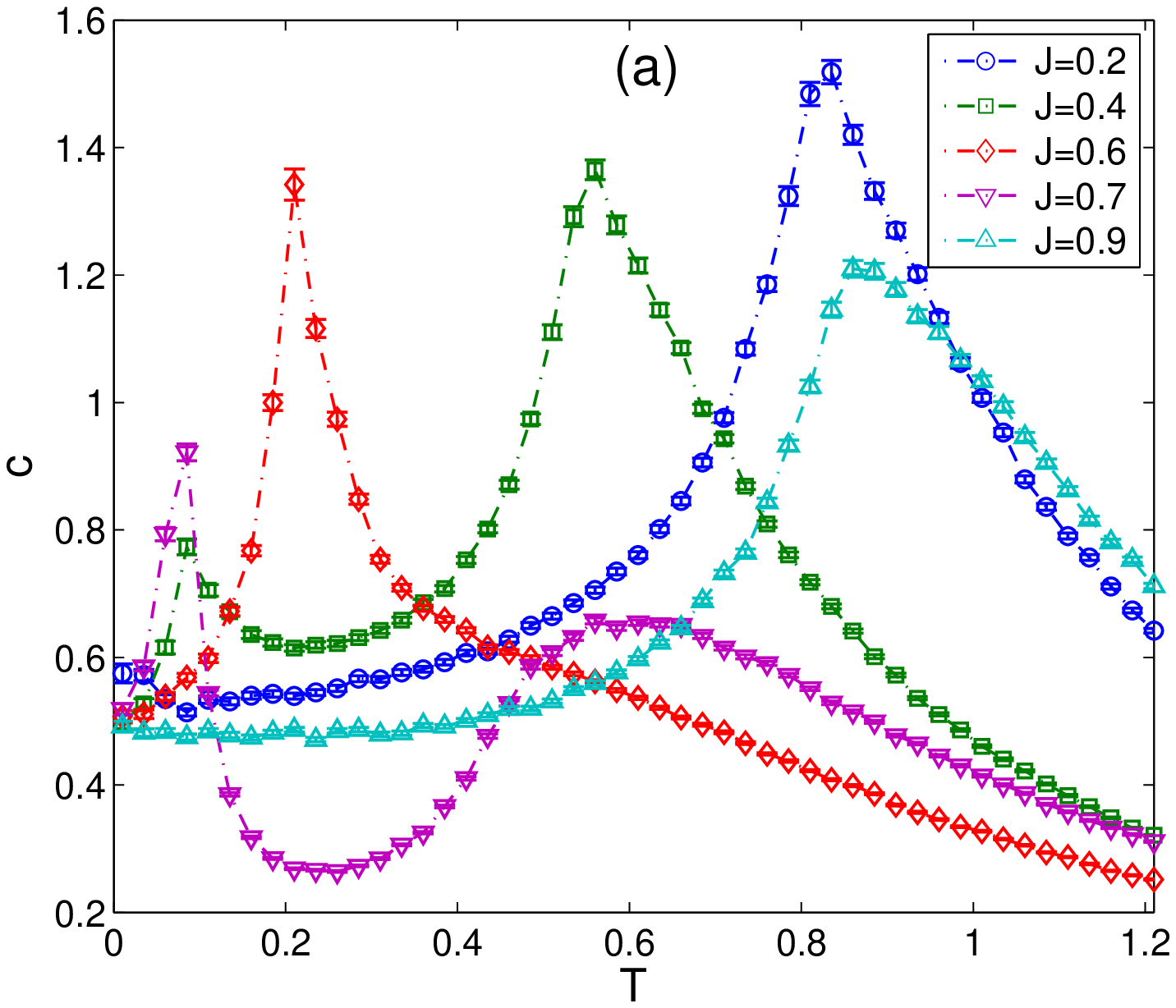}\label{fig:c-T}}
\subfigure{\includegraphics[scale=0.48,clip]{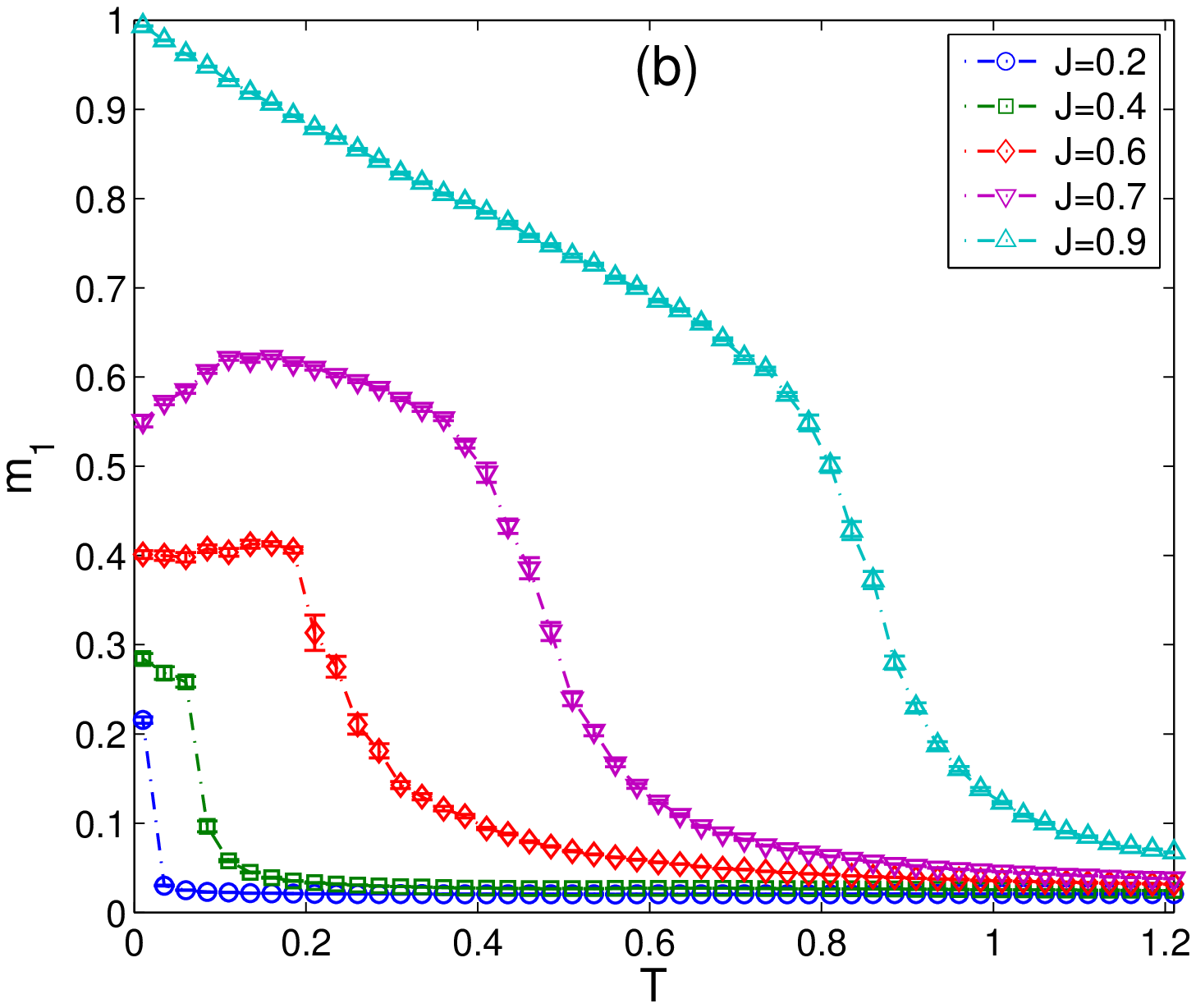}\label{fig:m-T}}\\
\subfigure{\includegraphics[scale=0.48,clip]{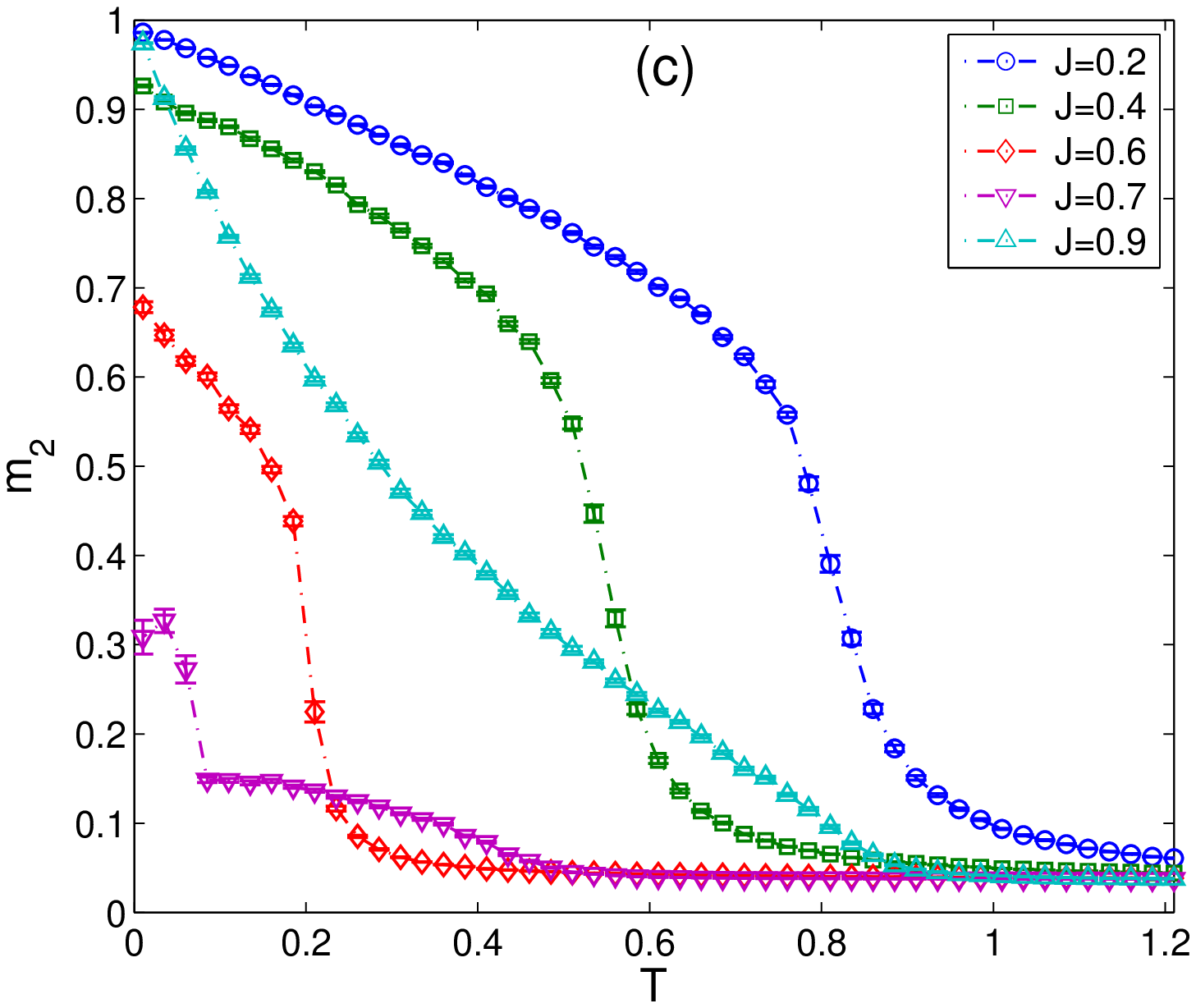}\label{fig:q-T}}
\subfigure{\includegraphics[scale=0.48,clip]{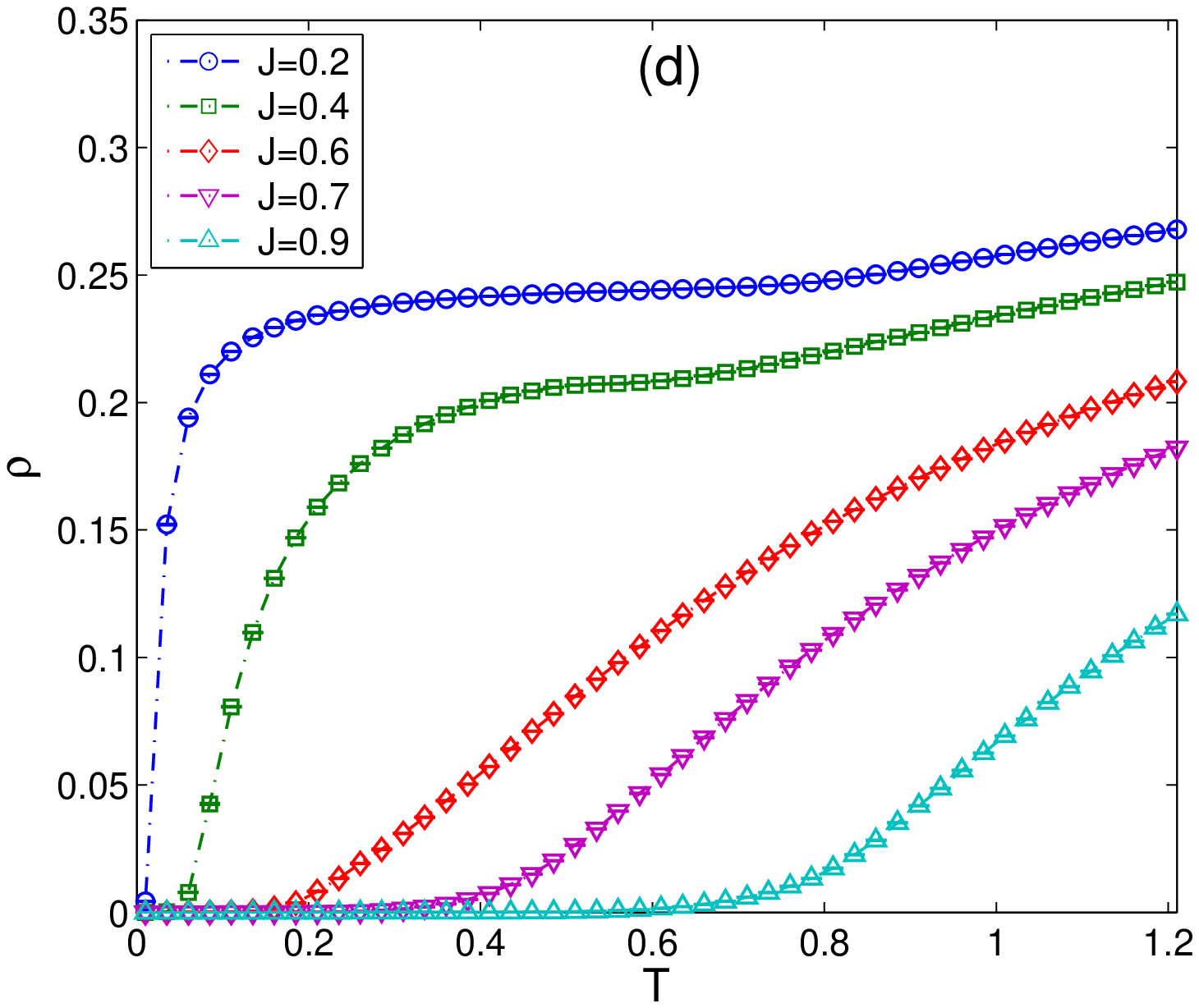}\label{fig:nv-T_v2}}
\caption{(Color online) Temperature dependencies of (a) the specific heat $c$, (b) the magnetization $m_1$, (c) the nematic order parameter $m_2$, and (d) the vortex density $\rho$, for various values of $J$.}\label{fig:therm_fun}
\end{figure}

Temperature dependencies of various functions, presented in Fig.~\ref{fig:therm_fun} for different values of $J$, provide insight into thermodynamic behavior of the system at finite temperatures. The specific heat curves display one apparent anomaly (maximum) above $J=0.8$ and two anomalies below $J=0.8$ (except for $J \approx 0.6$, as discussed below), indicating the occurrence of one and two phase transitions, respectively. 

In order to identify the nature of the ordering in the respective phases in the panels~\ref{fig:m-T} and~\ref{fig:q-T} we present the magnetic, $m_1$, and nematic, $m_2$, order parameters. They demonstrate that the magnetic phase appears at small values of $J$ but persists only at very low temperatures. As $J$ increases it gradually extends to higher temperatures. One can also notice that for $J < 0.8$ the magnetic order parameter fails to reach the saturation value of one, as it is in the cases of $J > 0.8$. Nevertheless, this behavior is expected based on the ground-state analysis presented above, which predicted the CFM phase for $J < 0.8$ and FM phase for $J > 0.8$.

\begin{figure}[t!]
\centering
\includegraphics[scale=0.6]{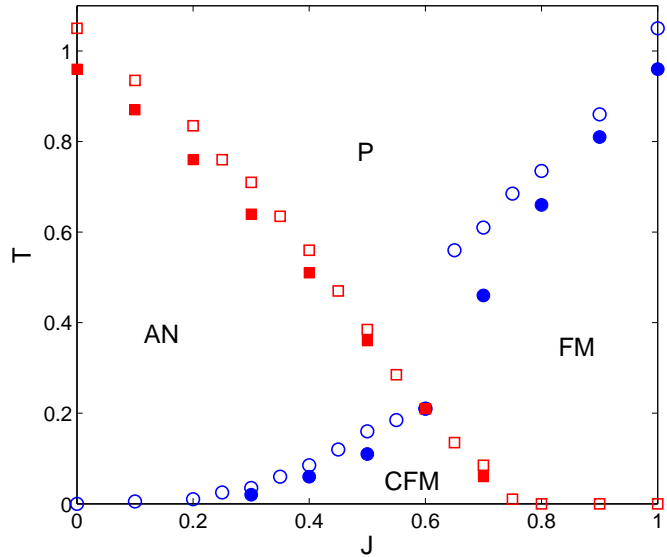}
\caption{(Color online) Phase diagram in $J$-$T$ plane. FM, AN, CFM, and P denote respectively the ferromagnetic, antinematic, canted ferromagnetic, and paramagnetic phases. Empty (filled) symbols represent transition points determined from the specific heat peaks (correlation function analysis).}\label{fig:PD}
\end{figure}

On the other hand, the nematic order parameter indicates the presence of the AN phase spanning to relatively high temperatures already at small values of $J$, where the FM ordering is absent. Then the extent of the AN phase is decreased by increasing $J$ up to $J \approx 0.6$, at which both $m_1$ and $m_2$ vanish at about the same temperature. Within $0.6 \lesssim J \lesssim 0.8$ the AN phase continues to shrink albeit $m_2$ remains finite for some temperature range even above the transition line, owing to the fact that the transition is not to the paramagnetic (P) but the FM phase. In Fig.~\ref{fig:nv-T_v2} we can observe an anomalous increase of the vortex density related to their unbinding at the transition from the FM state. The respective transition temperatures, estimated from the specific heat maxima, and the character of the respective phases, identified from the magnetic and nematic order parameters, summarized in a rough phase diagram, are presented in Fig.~\ref{fig:PD} (empty symbols).

We note that the order-disorder transition temperatures estimated from the specific heat peaks positions tend to overestimate the true values. One possibility of obtaining more reliable results is based on the helicity modulus $\Upsilon$ - a true order parameter that exhibits a universal jump at the transition temperature $T_{\rm BKT}$ from a finite value to zero in the disordered regime. In the standard $XY$ model $T_{\rm BKT}$ can be determined from the condition $\Upsilon(T_{\rm BKT}) = 2T_{\rm BKT}/\pi\nu^2$, where $\nu$ is the vorticity. In the generalized $XY$ model with the mixed vorticities, like ours with $\nu_1=1$ and $\nu_2=1/2$, some previous studies interpolated between the two pure cases by applying the condition $\Upsilon(T_{\rm BKT}) = 2 T_{\rm BKT}(J/\pi\nu_1^{2}+(1-J)/\pi\nu_2^{2})$~\cite{qi13,park08,qin09}. More recently such an approach has been questioned and an alternative approach, which instead of the helicity jump relies on scale invariance of the corresponding correlation function below the transition temperature, has been proposed~\cite{hubs13}. 

In the approach based on the correlation function analysis one essentially needs to find a temperature $T_{\rm BKT}$ separating two regimes: the power-law for $T<T_{\rm BKT}$ from the exponential one at $T>T_{\rm BKT}$. This can be accomplished by assessing the dependence given by Eq.~(\ref{g_FSS}) and finding the point at which the algebraic dependence ceases to be valid. Before doing so, let us focus more on the respective correlation functions $g_1$ and $g_2$.

\begin{figure}[t!]
\centering
\subfigure{\includegraphics[scale=0.48,clip]{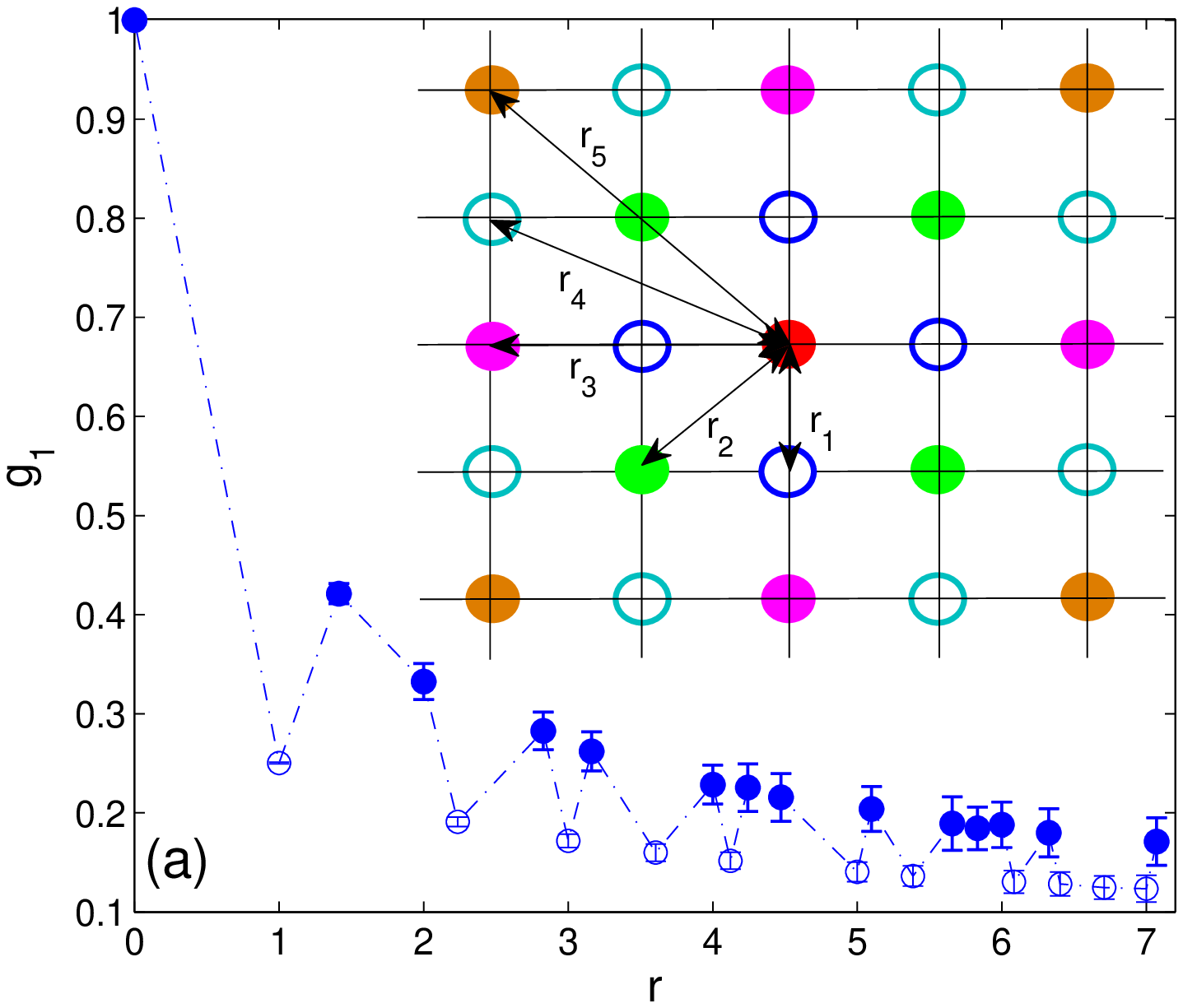}\label{fig:g1-r_J05_latt}}
\subfigure{\includegraphics[scale=0.48,clip]{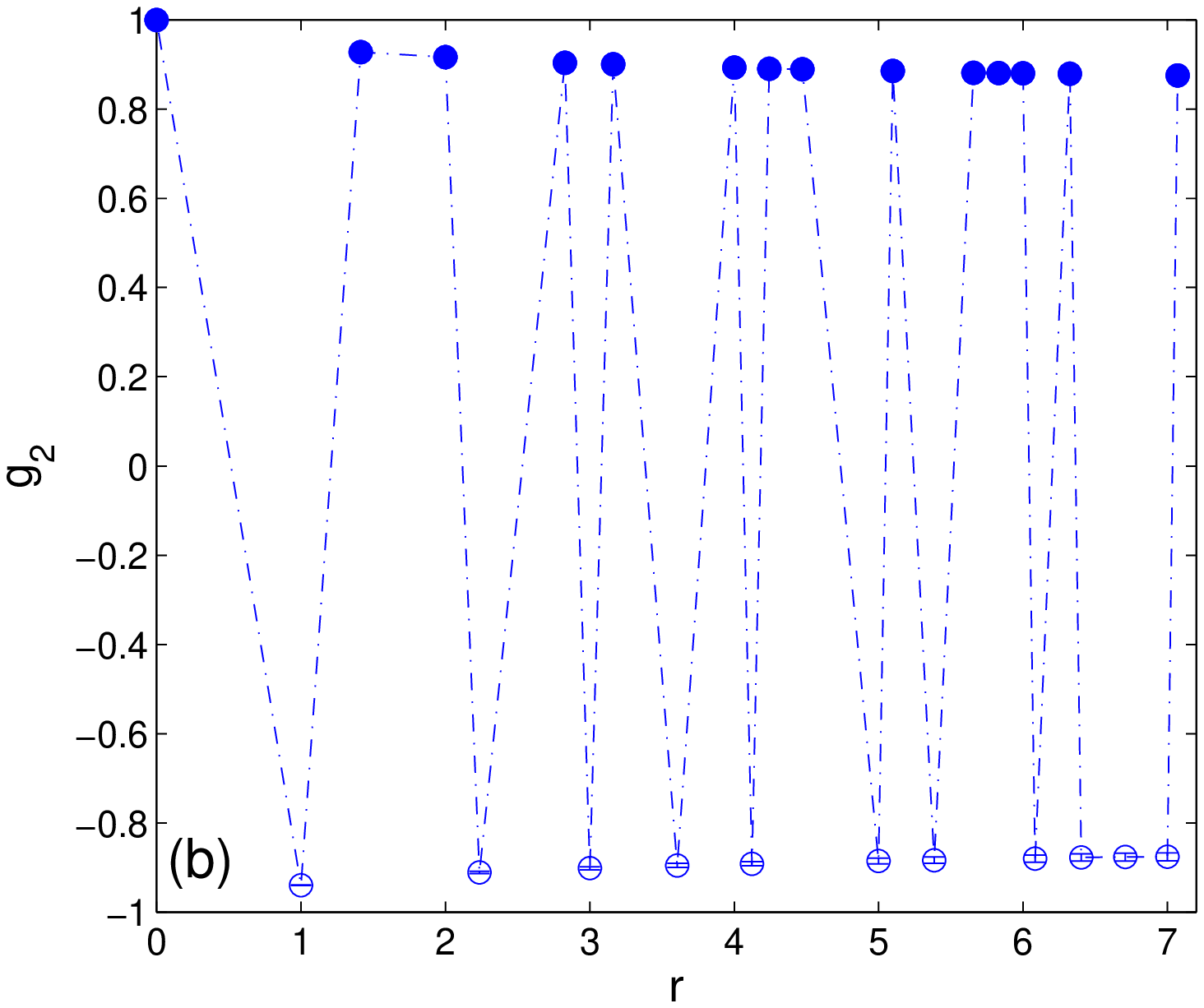}\label{fig:g2-r_J05}}
\caption{(Color online) Correlation functions (a) $g_1$ and (b) $g_2$, for $J=0.5$ and $T=0.01$. The inset in (a) schematically depicts the first five distance lags.}\label{fig:corr_fun_1}
\end{figure}

Due to the conflicting exchange interactions between nearest-neighbor spins belonging to different sublattices it is reasonable to assume that, at least within the CFM phase, the correlations between spins belonging to the same sublattice will be different from the correlations between spins belonging to different sublattices. Fig.~\ref{fig:g1-r_J05_latt} illustrates the situation for $J=0.5$ and $T=0.01$. The inset schematically shows a central spin (filled red circle) and spins in its near neighborhood at distance lags $r_1<r_2<\cdots<r_5$. From the correlation function $g_1(r)$ it is apparent that the correlations at the lags $r_2,r_3$ and $r_5$, i.e., between spins belonging to the same sublattice (filled circles) are larger than those at the lags $r_1$ and $r_4$, i.e., between spins belonging to different sublattices (empty circles). The difference is even much more striking in $g_2(r)$ (see Fig.~\ref{fig:g2-r_J05}), in which the quadrupoles belonging to the same (different) sublattice are strongly correlated (anticorrelated).

\begin{figure}[t!]
\centering
\subfigure{\includegraphics[scale=0.48,clip]{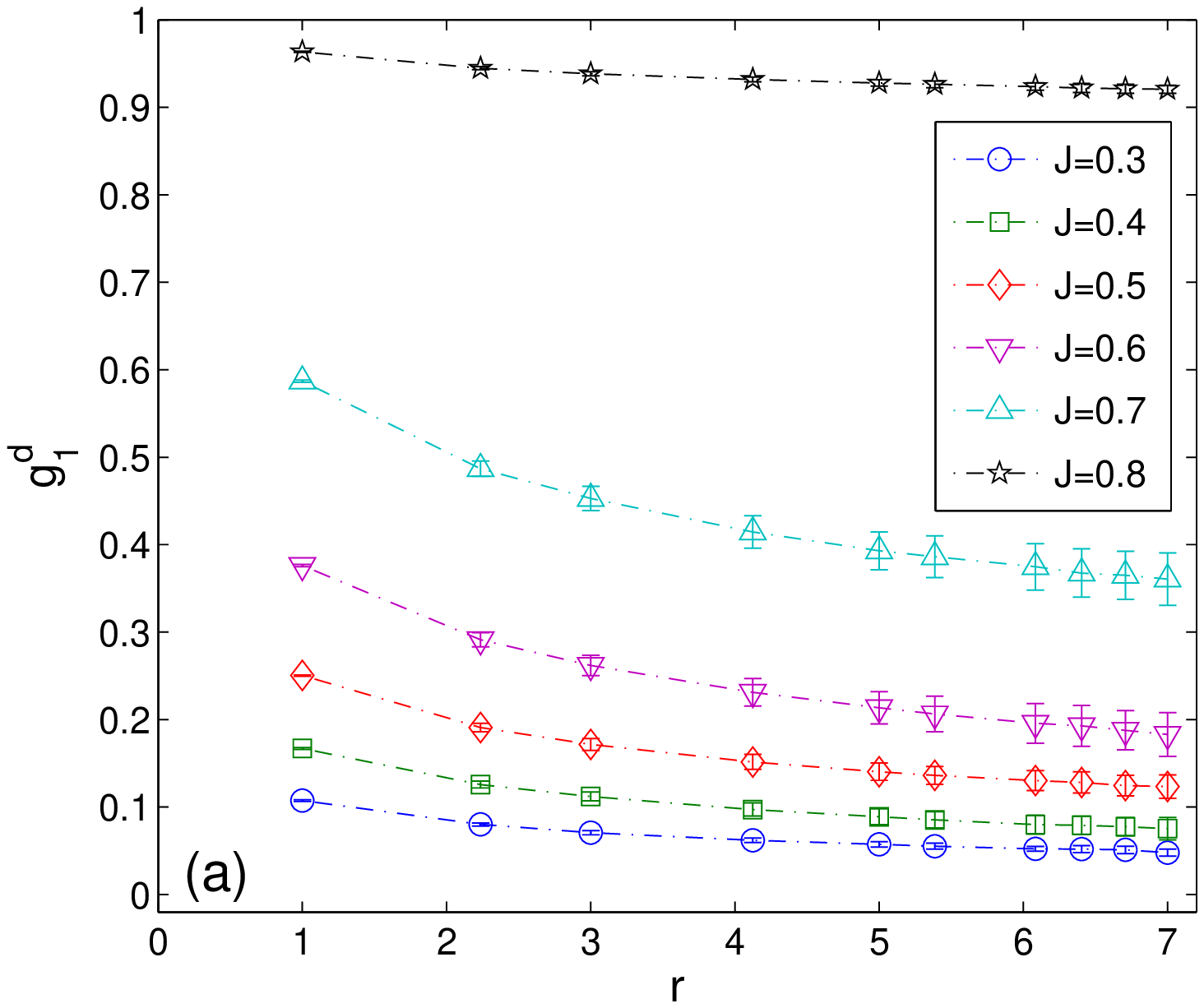}\label{fig:g1_1st-r}}
\subfigure{\includegraphics[scale=0.48,clip]{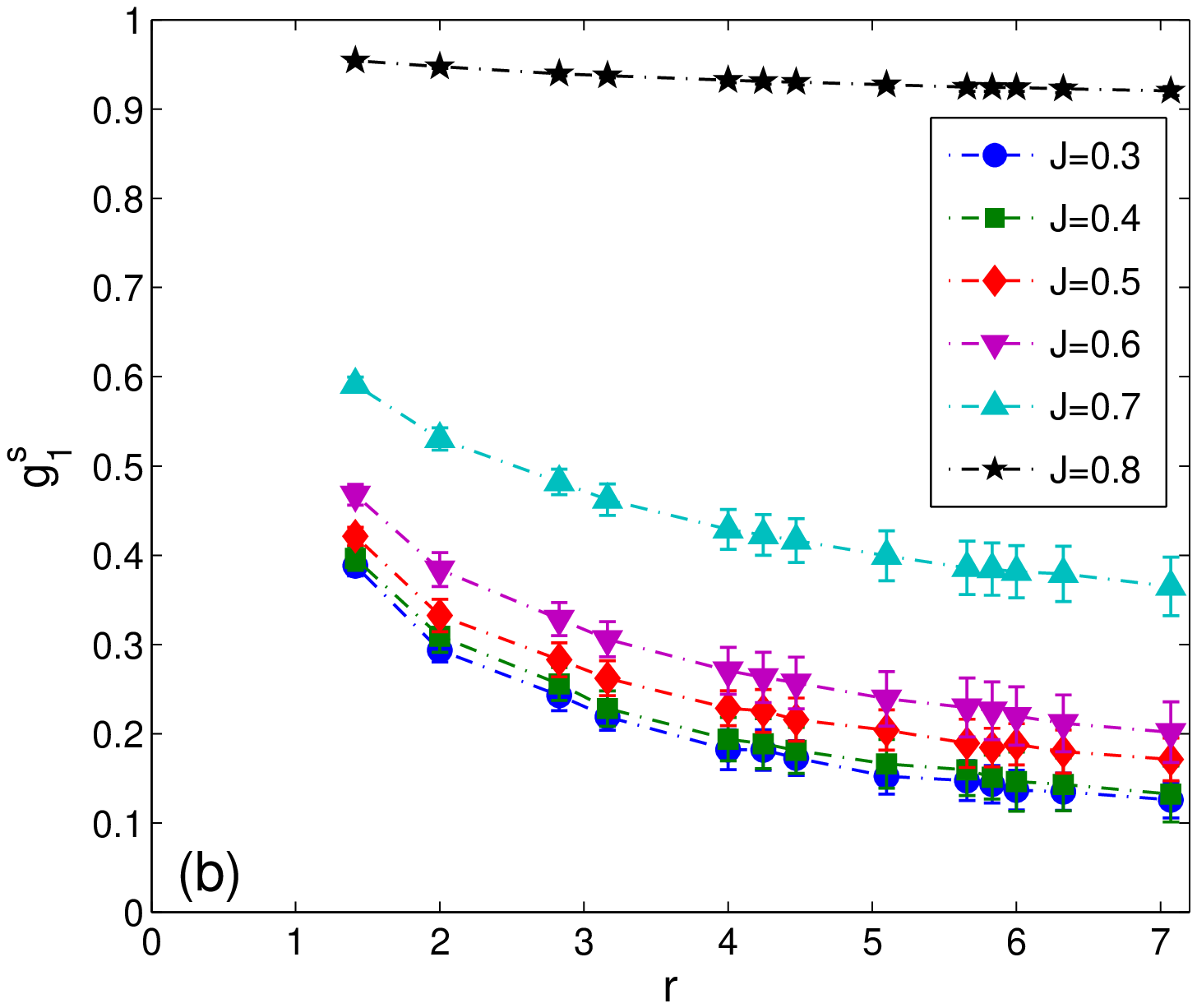}\label{fig:g1_2nd-r}}
\caption{(Color online) Correlation functions (a) $g_1^{d}$ and (b) $g_1^{s}$, obtained for different values of $J$ at $T=0.01$.}\label{fig:corr_fun_2}
\end{figure}

Therefore, in the following we split the correlation function $g_1$ into $g_1^{d}$, which includes the lags $r_1,r_4,\cdots$ between spins belonging to different sublattices and $g_1^{s}$, which includes the lags $r_2,r_3,\cdots$ between spins  belonging to the same sublattice. Their distance dependencies for various values of $J$ at $T=0.01$ are plotted in Fig.~\ref{fig:corr_fun_2}. While the differences between the two are barely noticeable at larger $J \lesssim 0.8$ they become much more pronounced at smaller values of $J$. Nevertheless, at sufficiently low temperatures they all follow power-law behavior with the exponent varying with both $J$ and $T$.

\begin{figure}[t!]
\centering
\subfigure{\includegraphics[scale=0.48,clip]{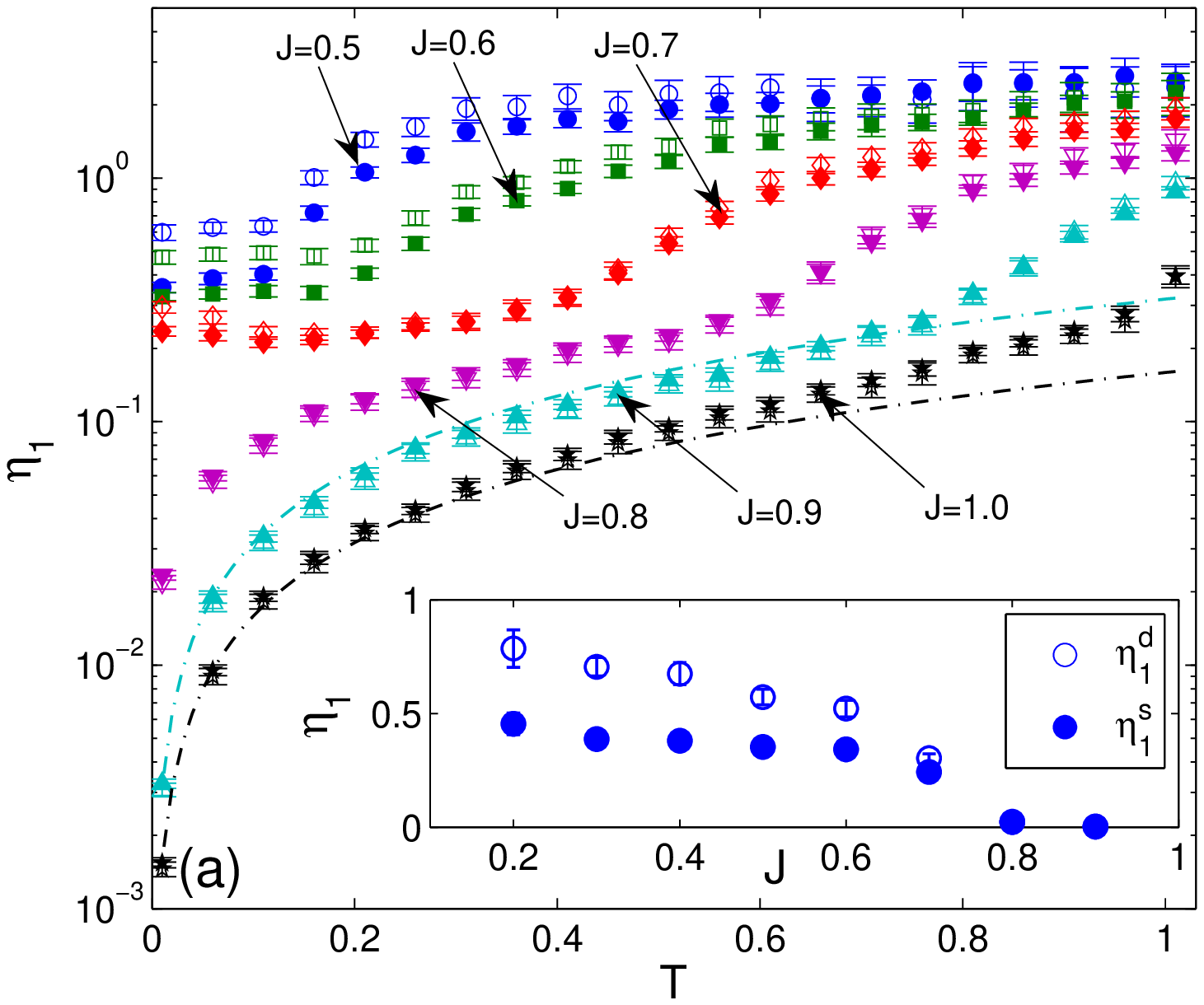}\label{fig:CF1-T_J05-1}}
\subfigure{\includegraphics[scale=0.48,clip]{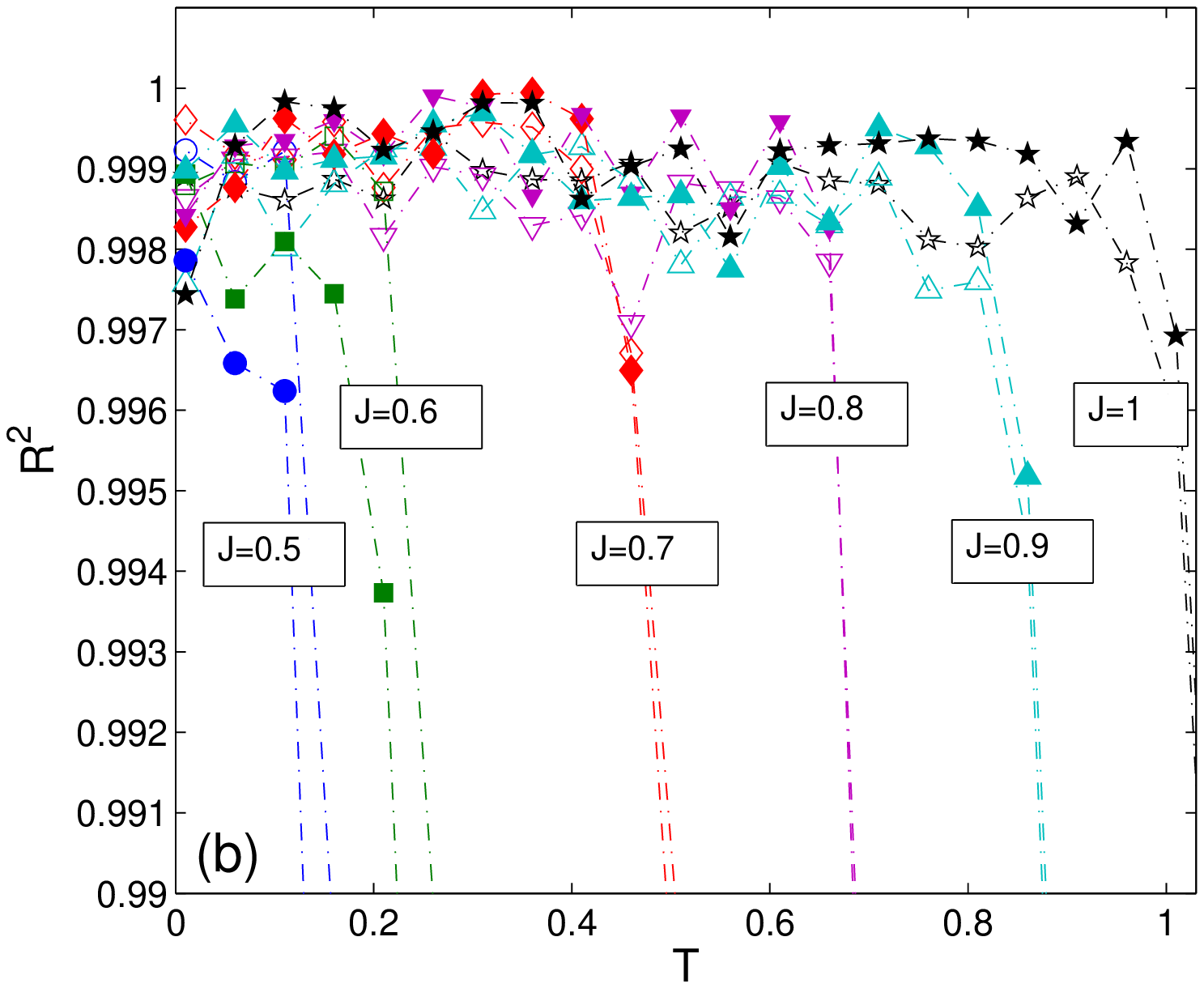}\label{fig:R2CF1-T_J05-1}}
\caption{(Color online) (a) Temperature dependencies of the critical exponents $\eta_1^{d}$ (empty symbols) and $\eta_1^{s}$ (filled symbols), obtained for different values of $J$, and (b) the corresponding adjusted coefficients of determination $R^2$. The dash-dotted curves in (a) for $J=0.9$ and 1 correspond to the spin-wave approximation.}\label{fig:eta}
\end{figure}

In Fig.~\ref{fig:CF1-T_J05-1}, we show temperature dependencies of the correlation function critical exponents $\eta_1^{d}$ (empty symbols) and $\eta_1^{s}$ (filled symbols) for various values of $J$. In line with the above comments the differences between the $\eta_1^{d}(T)$ and $\eta_1^{s}(T)$ curves are almost negligible for $J>0.7$, while $\eta_1^{d}(T) \geq \eta_1^{s}(T)$ for $J \leq 0.7$. The curves for $J=0.9$ and 1 are supplemented by the spin-wave approximation for $\eta_1$ (not distinguishing $\eta_1^{d}$ and $\eta_1^{s}$), which is applicable for the present generalized model if $J>0.8$ and acquires the form $\eta_{1}^{eff} = T/(2 \pi J^{eff})$, where the effective coupling $J^{eff} = 5J - 4$~\cite{zuko18}. The inset demonstrates the behavior of the two quantities as a function of $J$ for the fixed temperature $T=0.01$.

In the low-temperature limit of $T \to 0$ both the MC simulation as well as the spin-wave approximation indicate that for $J>0.8$ the ground state should correspond to $\eta_1^{d}=\eta_1^{s}=\eta_1=0$, i.e., the long-range ordering. On the other hand, for $J<0.8$ the MC simulation results suggest that $\eta_1^{d} > \eta_1^{s} > 0$ even in the ground state, i.e., the system remains in the QLRO CFM state.

At high temperatures all the curves appear to approach the limiting value of $\eta \approx 2$, which corresponds to the exponential decay of the correlation function, i.e., the paramagnetic state. However, the crossover between the two regimes is smeared (supposedly by a limited lattice size) and, thus, the temperatures at which the respective curves $\eta_1(T)$ reach the value of two overestimate the true transition temperatures. Better estimates can be obtained by monitoring of the adjusted coefficient of determination $R^2$~\cite{theil61}, presented in Fig.~\ref{fig:R2CF1-T_J05-1}, as a measure of goodness of the linear fit on a log-log scale expected in the algebraic phase. The latter corresponds to the values of $R^2 \approx 1$ and their sudden drop to lower values indicates deterioration of the linear fit due to the onset of the exponential dependence. 

Thus the temperatures at which this qualitative change in the $g_1$ function occurs can be considered as estimates of the transition temperatures at the onset of the FM phase. Analogically, a similar analysis of the $g_2$ function will provide us with the estimates of the transition temperatures to the AN phase. The phase diagram obtained by the above correlation function analysis is shown by the filled symbols in Fig.~\ref{fig:PD}. As already mentioned above, compared with the transition temperatures estimated from the specific heat maxima, the correlation function analysis gives lower values for any $J$, except $J \approx 0.6$ corresponding to the crossing point of the FM and AN branches at which both estimates are about the same.

\begin{figure}[t!]
\centering
\subfigure{\includegraphics[scale=0.48]{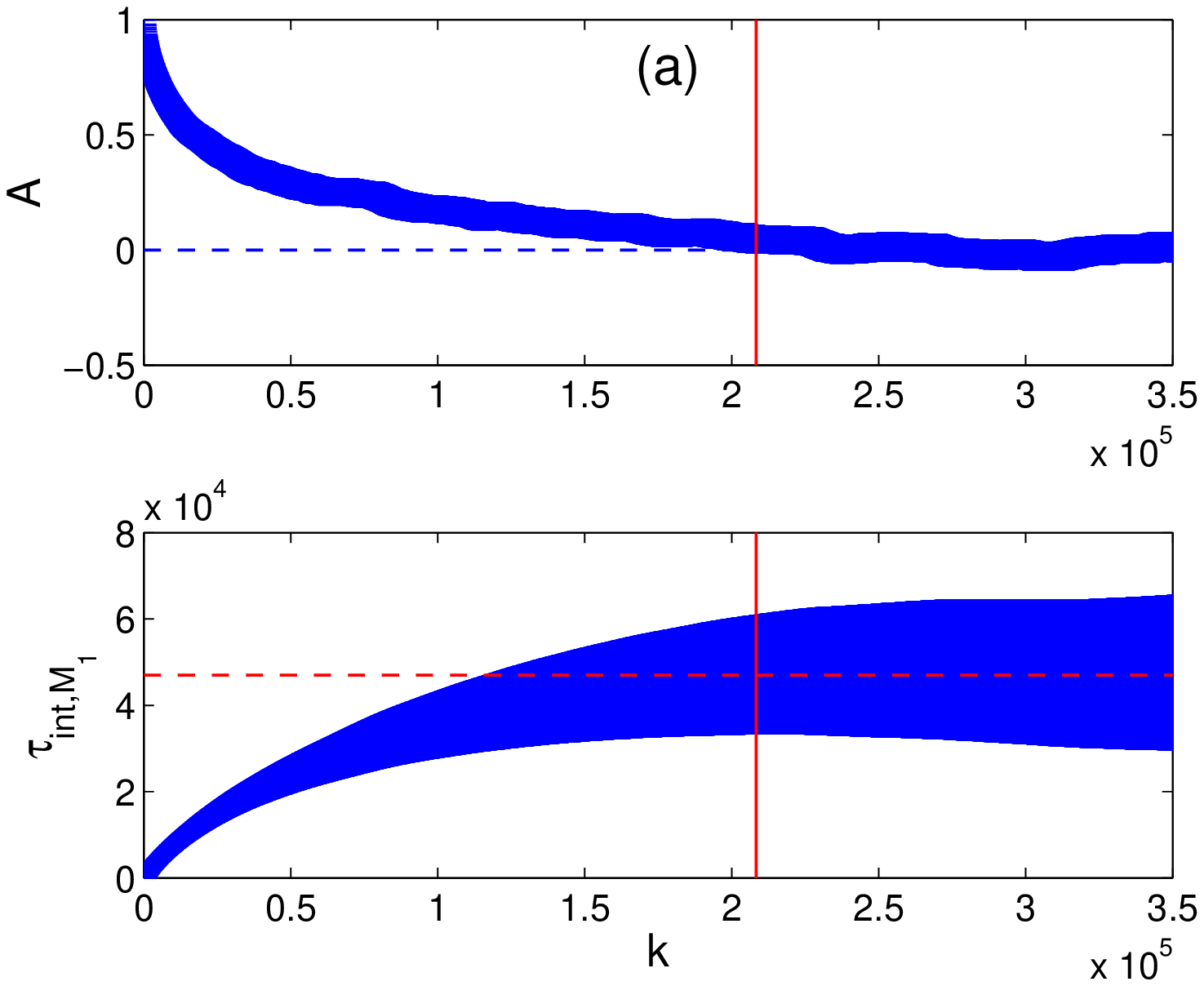}\label{fig:tau_int_J_0_4_m}}
\subfigure{\includegraphics[scale=0.48]{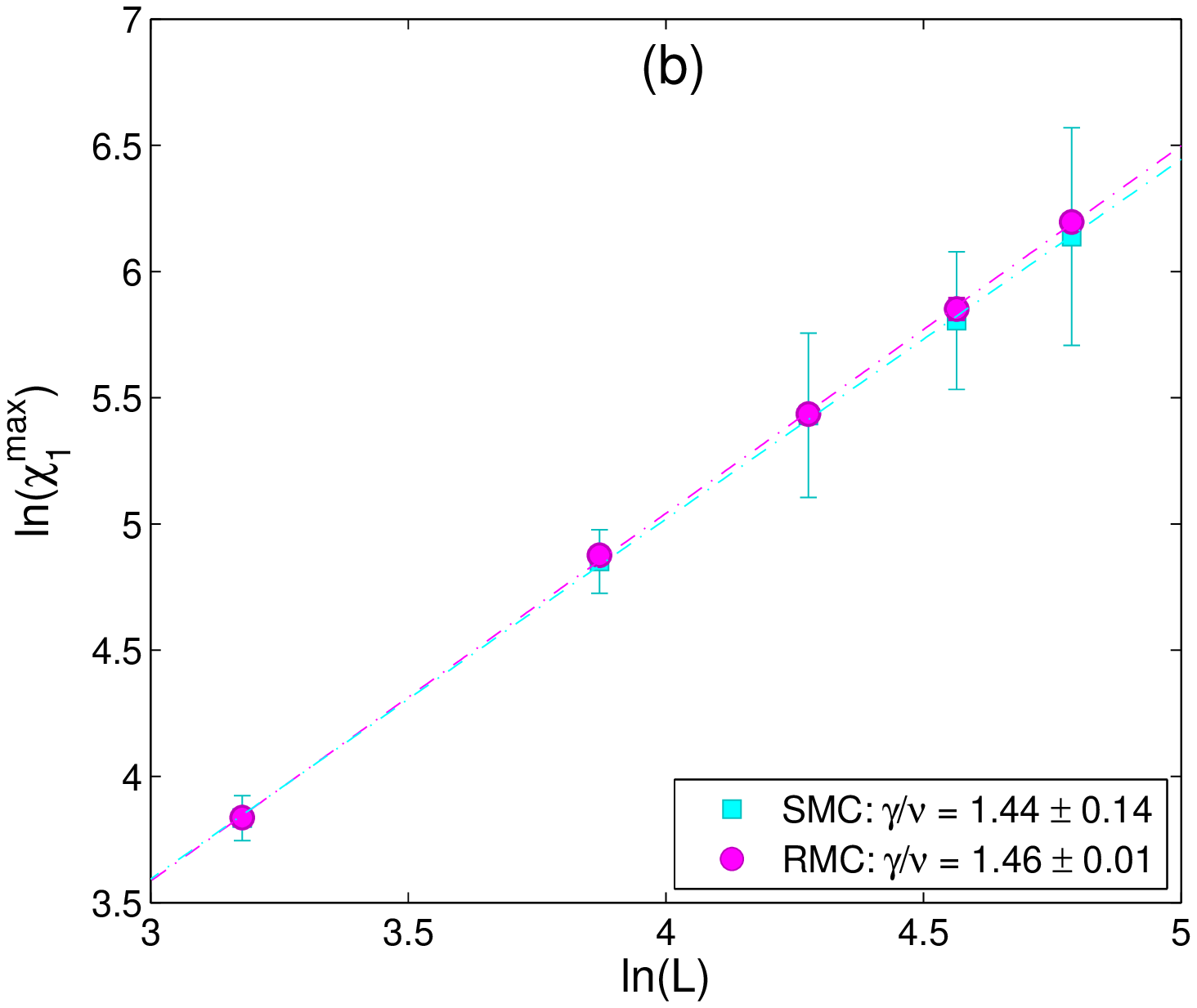}\label{fig:fss_J_0_4_m}}
\caption{(Color online) (a) Normalized autocorrelation function, $A(k)$, and the integrated autocorrelation time of the magnetization, $\tau_{int,M_1}$, close to the transition point at $T=0.0648$, for $L=120$ and $J=0.4$. (b) The FSS of the magnetic susceptibility at the AN-CFM phase boundary, for $J=0.4$. The cyan squares correspond to data from 10-15 independent standard MC (SMC) runs and the magenta circles from the reweighting (RMC) method.}\label{fig:J_0_4_m}
\end{figure}

Finally, we focus on the character of the transition between the AN and CFM phases. In both the non-frustrated $XY$ model with $J_1>0$ and $J_2>0$~\cite{kors85,lee85,carp89,hubs13} as well as the frustrated one with $J_1<0$ and $J_2<0$ on a triangular  lattice~\cite{park08} the transition between the nematic and magnetic phases was confirmed to belong to the Ising universality class. In order to verify if the same scenario also applies in the present system, we performed the FSS analysis at the AN-CFM branch of the phase diagram for $J=0.4$. We note that the transition occurs at rather low temperature and, therefore, care should be taken to properly handle potential equilibration problems and long autocorrelation times, particularly for large system sizes. As demonstrated in Fig.~\ref{fig:tau_int_J_0_4_m}, which for the largest considered size $L=120$ shows the normalized autocorrelation function $A$ of the magnetization as a function of the time lag $k$ (upper panel) and the integrated autocorrelation time $\tau_{int,M_1}$ (lower panel), the relatively large value of the latter $\tau_{int,M_1} \approx (4.7 \pm 1.2) \times 10^4$ considerably reduces the effective sample size. To avoid any spurious results we first performed the FSS analysis based on 10-15 independent standard MC simulation runs for each $L$ in a small range of temperatures around the expected transition point with the fine resolution of $\Delta T=0.001$. Subsequently those were supplemented by much longer runs using $10^7$ MCS at just one temperature close to the pseudo-transition point (different for each $L$) followed by the reweighting method to determine extrema of the relevant quantities.

In particular, the FSS analysis of maxima of the magnetic susceptibility is presented in Fig.~\ref{fig:fss_J_0_4_m}. Due to the reasons mentioned above the plot of the FSS relation (\ref{fss_chi}) using data from the standard MC (SMC) simulations involves relatively large error bars for the increasing $L$. Nevertheless, the critical exponents ratio $\gamma/\nu = 1.44 \pm 0.14$ corresponds rather well with that obtained from the reweighting (RMC) method ($\gamma/\nu = 1.46 \pm 0.01$) and both clearly differ from the Ising universality, for which the expected value is $\gamma_{I}/\nu_{I}=7/4$. We assume that the deviation from the Ising universal behavior is caused by the competition between the two types of couplings, resulting in the frustrated (canted) structure of the CFM phase. We note that failure of the FSS with the Ising critical exponents at the nematic-magnetic phase boundary was also reported in the frustrated $XY$ model on a square lattice and was ascribed to the frustration present in the system~\cite{qin09}.

\section{Summary}

We studied critical properties of the generalized $XY$ model with the ferromagnetic (FM) $J_1 \equiv J \in (0,1)$ and antinematic (AN) $J_2 = J-1 < 0$ nearest-neighbor interactions on a square lattice for a varying $J$. We found that the ground state of the system is ferromagnetic (FM) for sufficiently large values of the FM couplings, namely $J \in [0.8,1)$. For $J \in (0,0.8)$ the ground state corresponds to an intricate canted ferromagnetic (CFM) state, resulting from the competition between the collinear FM and noncollinear AN ordering. At finite temperatures as $T$ is lowered there are two successive phase transitions within $J \in (0,0.6)$: first from the paramagnetic (P) to the AN phase at intermediate temperatures followed by another one to the CFM phase at very low temperatures. There are two successive phase transitions also for $J \in (0.6,0.8)$. Within this range the transition from the P phase is to the FM phase followed by the FM-CFM transition at lower temperatures. Close to $J \approx 0.6$, at which the P-AN and P-FM phase transition boundaries cross,  there appears to exist only one transition from the P directly to the CFM phase. The CFM phase is absent within $J \in [0.8,1)$ and, therefore, there is also only one phase transition from the P to the FM phase. The peculiar low-temperature CFM phase is characterized by highly degenerate states in which neighboring spins that belong to different sublattices are canted by a non-universal angle $\pm \phi_{GS}(J)$. Spin-pair correlation functions are different for spins belonging to the same and different sublattices but both retain the power-law decaying character down to zero temperatures. Nevertheless, the critical exponent $\eta$ in the CFM phase acquires much larger values, i.e., the correlation function decays much faster, than in the FM phase for the same temperature. The AN-CFM phase transition does not comply with the Ising universality class.

\begin{acknowledgments}
This work was supported by the Scientific Grant Agency of Ministry of Education of Slovak Republic (Grant No. 1/0531/19) and the scientific grants of Slovak Research
and Development Agency provided under contract No. APVV-16-0186.
\end{acknowledgments}

\end{document}